\documentclass[preprint,aps,amsmath,superscriptaddress,nofootinbib,tightenlines,showpacs]{revtex4}
\usepackage{amsmath}
\usepackage{amsfonts}
\usepackage{amssymb}
\usepackage{latexsym}
\usepackage{graphicx}
\usepackage{bm}
\usepackage{color}
\usepackage{epsfig}
\usepackage[english]{babel}

\begin{document}
\title{Membrane paradigm of black holes in Chern-Simons modified gravity}
\author{Tian-Yi Zhao}
\email[Electronic address: ]{zhaotianyi5566@foxmail.com}
\affiliation{Department of Physics, East China Normal University,\\
No. 500 Dongchuan Rd., Shanghai 200241, China}
\author{Towe Wang}
\email[Electronic address: ]{twang@phy.ecnu.edu.cn}
\affiliation{Department of Physics, East China Normal University,\\
No. 500 Dongchuan Rd., Shanghai 200241, China}
\affiliation{Shanghai Key Laboratory of Particle Physics and Cosmology,\\
Department of Physics and Astronomy, Shanghai Jiao Tong University,\\
No. 800 Dongchuan RD., Shanghai 200240, China\\ \vspace{0.2cm}}
\date{\today\\ \vspace{1cm}}
\begin{abstract}
The membrane paradigm of black hole is studied in the Chern-Simons modified gravity. Derived with the action principle a la Parikh-Wilczek, the stress tensor of membrane manifests a rich structure arising from the Chern-Simons term. The membrane stress tensor, if related to the bulk stress tensor in a special form, obeys the low-dimensional fluid continuity equation and the Navier-Stokes equation. This paradigm is applied to spherically symmetric static geometries, and in particular, the Schwarzschild black hole, which is a solution of a large class of dynamical Chern-Simons gravity.
\end{abstract}


\maketitle




\section{Introduction}\label{sect-intro}
The membrane paradigm, developed three decades ago on the basis of field equations of Einstein gravity \cite{Damour:1978cg,Damour79,Damour82,Price:1986yy,Thorne86}, is a valuable formalism for studying the black hole hydrodynamics. On the event horizon of black hole, the gravitational equations resemble the low-dimensional fluid continuity equation and the Navier-Stokes equation. The fluid quantities, such as density, pressure, shear and expansion, can be read from the stretched horizon up to a renormalization parameter.

One and a half decades ago, in Parikh and Wilczek's work \cite{Parikh:1997ma}, the stress tensor of low-dimensional fluid was derived from an action with a surface term on the stretched horizon. This method is powerful for the black hole membrane paradigm. Interesting related results can be also found in \cite{Ashtekar:2000hw,Booth:2001gx,Padmanabhan:2010rp,Jaramillo:2013rda} as a partial list. Along the Parikh-Wilczek method, the membrane paradigm has since then been studied in $f(R)$, Gauss-Bonnet and Love-Lock gravity theories \cite{Chatterjee:2010gp,Jacobson:2011dz,Kolekar:2011gg}. Very recently, this method was utilized in analyzing a new example of membrane paradigm: an oblique membrane paradigm for cosmological horizon \cite{Wang:2014aty}.

The Chern-Simons modified gravity \cite{Jackiw:2003pm,Alexander:2009tp} is an effective extension of general relativity that captures leading-order gravitational parity violation. It continues to support the Schwarzschild solution in a large class of extended theory \cite{Jackiw:2003pm,Grumiller:2007rv}. In this article, following the Parikh-Wilczek method \cite{Parikh:1997ma}, we investigate the membrane paradigm of black holes in Chern-Simons modified gravity, firstly in a general framework and then focusing on spherically symmetric static geometries with an emphasis on the Schwarzschild black hole.\footnote{As a partial list, see \cite{Lemos:2011ic,Delsate:2014hba} and references therein for recent development along this direction.}

This article is outlined as follows. In section \ref{sect-CS}, we will set up the general formalism for membrane paradigm in Chern-Simons gravity, including the membrane stress tensor from action principle, the relation between membrane and bulk stress tensors, and fluid equations. In section \ref{sect-sphst}, restricted to spherical static spacetimes, we decompose the Riemann tensor and extrinsic curvature tensor with no more than eight parameters, and put our general set-up in a concrete form. In subsection \ref{subsect-mbrel}, we exactly prove the relation between membrane stress tensor and the bulk stress tensor, which is necessary for the derivation of fluid equations. The fluid quantities are worked out in subsection \ref{subsect-fluqn}, where we find the Hall viscosity and momentum density are nonzero, even though the H\'a\'{\j}i\v{c}ek field vanishes. The Schwarzschild spacetime is studied as a specific example in subsection \ref{subsect-sch}. In section \ref{sect-outl}, we will present some outlooks for future investigation along this line.

\section{Membrane paradigm in Chern-Simons modified gravity}\label{sect-CS}
\subsection{Geometric notations and conventions}\label{subsect-geom}
In this section, we will begin with the geometric set-up which is very helpful for understanding the membrane paradigm. Recently these notations and conventions have been widely utilized in the black hole membrane paradigm in Einstein, $f(R)$, Gauss-Bonnet and Love-Lock gravity theories \cite{Parikh:1997ma,Chatterjee:2010gp,Jacobson:2011dz,Kolekar:2011gg}. They should be equally useful for exploring the same paradigm in the Chern-Simons modified gravity.

The black hole event horizon, $\mathcal{H}$, is a 3-dimensional null hypersurface with a null geodesic generator $l^{a}$. At the event horizon, the geodesic equation is $l^{a}\nabla_{a}l^{b}=g_{\mathcal{H}}l^{b}$, in which $g_{\mathcal{H}}$ is a nonaffine coefficient. In a stationary spacetime, $l^{a}$ is the null limit of the timelike Killing vector and $g_{\mathcal{H}}$ can be regarded as the surface gravity at the horizon \cite{Chatterjee:2010gp}. We can construct the membrane paradigm on the event horizon, but it is more convenient to formulate a timelike stretched horizon, which is outside $\mathcal{H}$ and very close to it. The stretched horizon has a natural advantage that the metric is nondegenerate thereon. The stretched horizon, notated by $\mathcal{S}$, is generated by the timelike congruence $u^{a}$ and has a spacelike normal vector $n^{a}$. Vectors $u^{a}$ and $n^{a}$ are orthogonal to each other and normalized to unity
\begin{equation}\label{norm}
n_{a}u^{a}=0,~~~~n^{a}n_{a}=1,~~~~u^{a}u_{a}=-1.
\end{equation}
The nondegenerate metric of the stretched horizon is given by
\begin{equation}\label{}
h_{ab}=g_{ab}-n_{a}n_{b}.
\end{equation}
There is a 2-dimensional spacelike cross section normal to $u^{a}$ with the metric
\begin{equation}\label{}
\gamma_{ab}=h_{ab}+u_{a}u_{b}.
\end{equation}
Then we can define the extrinsic curvature of stretched horizon as
\begin{equation}\label{}
K_{ab}=h_{b}^{~d}\nabla_{d}n_{a}=\nabla_{a}n_{b}.
\end{equation}

In the membrane paradigm, we introduce a parameter $\alpha$. When $\alpha\rightarrow 0$, the stretched horizon will tend to the event horizon, and $\alpha u^{a}\rightarrow l^{a}$, $\alpha n^{a}\rightarrow l^{a}$. At the same time, the components of $K_{ab}$ become
\begin{eqnarray}\label{Klim}
\nonumber\alpha u^{a}u^{b}K_{ab}&=&u^{a}u^{b}\nabla_{b}(\alpha n_{a})\rightarrow-g_{\mathcal{H}},\\
\alpha\gamma^{a}_{~A}\gamma^{b}_{~B}K_{ab}&=&\gamma^{a}_{~A}\gamma^{b}_{~B}\nabla_{b}(\alpha n_{a})\rightarrow k_{AB},
\end{eqnarray}
where $k_{AB}$ is the extrinsic curvature of the 2-dimensional spacelike section of the event horizon,
\begin{equation}\label{}
k_{AB}=\gamma^{a}_{~A}\gamma^{b}_{~B}\nabla_{b}l_{a}=\frac{1}{2}\mathcal{L}_{l}\gamma_{AB}.
\end{equation}
Here $\mathcal{L}_{l}$ is the Lie derivative along $l^{a}$. It proves convenient to decompose $k_{AB}$ into a traceless part and a trace,
\begin{equation}\label{}
k_{AB}=\sigma_{AB}+\frac{1}{2}\theta\gamma_{AB},
\end{equation}
where $\sigma_{AB}$ is the shear of $l_{a}$. Any traceless symmetric 2-dimensional tensor has two degrees of freedom and thus can be decomposed by $\sigma_{AB}$ and
\begin{equation}\label{sigtd}
\tilde{\sigma}^{AB}=\frac{1}{2}(\epsilon^{Adef}n_{d}u_{e}\sigma_{f}^{~B}+\epsilon^{Bdef}n_{d}u_{e}\sigma_{f}^{~A})
\end{equation}
Our conventions for $\tilde{\sigma}^{AB}$ is in accordance with references \cite{Saremi:2011ab,Jensen:2011xb}.

Throughout this paper, we will restrict to stationary spacetime with vanishing vorticity, namely
\begin{equation}\label{}
n^{a}\nabla_{a}n_{b}=0,~~~~\nabla_{a}n_{b}=\nabla_{b}n_{a},~~~~\gamma^{a}_{~A}\gamma^{b}_{~B}\nabla_{a}l_{b}=\gamma^{a}_{~A}\gamma^{b}_{~B}\nabla_{b}l_{a}.
\end{equation}
In this case, the limit $\alpha\rightarrow0$ gives \cite{wang:2016}
\begin{eqnarray}\label{Kdulim}
\nonumber K_{ab}&\rightarrow&\alpha^{-1}k_{ab}+\Omega_{a}u_{b}+\Omega_{b}u_{a}-\alpha^{-1}gu_{a}u_{b},\\
\nabla_{a}u_{b}&\rightarrow&\alpha^{-1}k_{ab}+\Omega_{a}n_{b}+\Omega_{b}n_{a}-\alpha^{-1}gu_{a}n_{b},
\end{eqnarray}
where the H\'a\'{\j}i\v{c}ek field
\begin{equation}\label{Hajek}
\Omega_{A}=\frac{1}{2}\gamma_{A}^{a}(n^{b}\nabla_{b}u_{a}-u^{b}\nabla_{b}n_{a}).
\end{equation}

\subsection{Membrane stress tensor from action principle}\label{subsect-tcs}
We will construct the membrane paradigm  in Chern-Simons modified gravity in a ($3+1$)-dimensional spacetime. When deriving the Euler-Lagrange equations at the event horizon, the action outside the horizon $S^{out}$ is not stationary unless we add a surface term $S^{surf}$. Therefore, the total action should be split as
\begin{equation}\label{}
S^{total}=(S^{in}-S^{surf})+(S^{surf}+S^{out}).
\end{equation}
Here $S^{surf}$ is the surface term, which represents the contribution of the stretched horizon. It satisfies $\delta S^{out}+\delta S^{surf}=0$. Without taking $S^{surf}$ into account, we cannot get the correct gravitational equations outside the horizon. In reference \cite{Parikh:1997ma}, it was noted for the first time that the contribution of the surface term can be encoded to the energy-momentum tensor of the $(2+1)$-dimensional fluid on the stretched horizon. Explicitly, the variation of the surface term is related to the $(2+1)$-dimensional energy-momentum tensor $t^{bd}$ by
\begin{equation}\label{vsurf}
\delta S^{surf}=\frac{1}{2}\int_{\mathcal{S}}d^{3}x\sqrt{h} t^{bd}\delta h_{bd}.
\end{equation}

The action of Chern-Simons modified gravity comprises the Einstein-Hilbert action and the Chern-Simons term. Incorporating boundary terms at spatial infinity, we can write down the action outside the event horizon as \cite{Grumiller:2008ie}
\begin{eqnarray}
\nonumber S^{out}&=&S_{EH}+S_{CS}+S_{GHY}+S_{bCS}+S_{\mathcal{F}}\\
\nonumber&=&\frac{1}{16\pi G}\int d^{4}x\sqrt{-g} [R+\frac{1}{4}\theta_{CS}^{~~~*}RR] \\
\nonumber&&+\frac{1}{8\pi G}\int_{\infty}d^{3}x\sqrt{h}(K+\frac{1}{2}\theta_{CS}n_{a}\epsilon^{abcd}K_{b}^{~e}\nabla_{c}K_{de})\\
&&+\frac{1}{16\pi G}\int_{\infty}d^{3}x\sqrt{h} \mathcal{F}(h_{ab}, \theta_{CS}),
\end{eqnarray}
where $S_{GHY}+S_{bCS}$, given by the next-to-last line, is the appropriate generalization of the Gibbons-Hawking term at infinity \cite{York:1972sj,Gibbons:1976ue}. For a variational principle to work in asymptotically flat spacetimes, other local counter terms may be necessary \cite{Mann:2005yr}. These counter terms, denoted by $S_{\mathcal{F}}$ here, are put to the last line. In the Chern-Simons term $S_{CS}$, the Chern-Pontryagin density is defined by
\begin{equation}\label{}
^{*}RR:={}^{*}R^{a~cd}_{~b}R^{b}_{~acd}=\frac{1}{2}\epsilon^{cdef}R^{a}_{~bef}R^{b}_{~acd}.
\end{equation}

The membrane paradigm in Einstein gravity is well-established. We will extend it to Chern-Simons modified gravity. For this purpose, we should calculate the $(2+1)$-dimensional energy-momentum tensor $t^{bd}$ according to equation \eqref{vsurf}. In the literature, this is always done in a circumbendibus, workable even if one does not know the explicit form of surface terms at the horizon and boundary terms at infinity. The point is that, on the stretched horizon, the boundary terms at infinity get irrelevant, and we can trade $\delta S^{surf}$ to $\delta S^{out}$ by $\delta S^{out}+\delta S^{surf}=0$. Along this logic, it is enough here to calculate $\delta S_{EH}+\delta S_{CS}$. In reference \cite{Parikh:1997ma}, the variation of $S_{EH}$ has been worked out already, yielding the partial result
\begin{equation}\label{teh}
t_{EH}^{bd}=\frac{1}{8\pi G} (Kh^{bd}-K^{bd}).
\end{equation}
Therefore, in this section, we will focus on $\delta S_{CS}$, and combine $t_{EH}^{bd}$ with $t_{CS}^{bd}$ in the end.

Because we do not care about the terms that are not total derivatives, or boundary terms that vanish on the stretched horizon, we use $\simeq$ to denote equivalence up to these terms. In this notation, with the help of identities
\begin{eqnarray}\label{}
\nonumber\delta R^{b}_{~acd}&=&\nabla_{c}\delta\Gamma^{b}_{~ad}-\nabla_{d}\delta\Gamma^{b}_{~ac},\\
\delta\Gamma^{b}_{~ac}&=&\frac{1}{2}g^{bd}(\nabla_{a}\delta g_{dc}+\nabla_{c}\delta g_{ad}-\nabla_{d}\delta g_{ac}),
\end{eqnarray}
we can get
\begin{eqnarray}
\nonumber\delta S_{CS}&\simeq&\frac{1}{64\pi G}\int d^{4}x\sqrt{-g}\theta_{CS}\delta(^{*}RR)\\
\nonumber&\simeq& \frac{1}{16\pi G}\int d^{4}x\sqrt{-g}\theta_{CS}^{~~~*}R^{a~cd}_{~b}\nabla_{c}(\delta\Gamma^{b}_{~ad})\\
\nonumber&\simeq& \frac{1}{32\pi G}\int d^{4}x\sqrt{-g}\theta_{CS}^{~~~*}R^{a~cd}_{~b}g^{be}\nabla_{c}(\nabla_{a}\delta g_{ed}+\nabla_{d}\delta g_{ae}-\nabla_{e}\delta g_{ad})  \\
&\simeq& \frac{1}{16\pi G}\int d^{4}x\sqrt{-g}\theta_{CS}^{~~~*}R^{abcd}\nabla_{c}\nabla_{a}\delta g_{bd}\\
\nonumber&\simeq&  \frac{1}{16\pi G}\int d^{4}x\sqrt{-g}\nabla_{c}(\theta_{CS}^{~~~*}R^{abcd}\nabla_{a}\delta g_{bd})-\frac{1}{16\pi G}\int d^{4}x\sqrt{-g}\nabla_{a}[\nabla_{c}(\theta_{CS}^{~~~*}R^{abcd})\delta g_{bd}].
\end{eqnarray}
Choosing $n^a$ to be outward-pointing, we can use the Gauss theorem to obtain
\begin{eqnarray}\label{vScs}
\nonumber\delta S_{CS}&\simeq&-\frac{1}{16\pi G}\int_{\mathcal{S}}d^{3}x\sqrt{-h}n_{c}\theta_{CS}^{~~~*}R^{abcd}\nabla_{a}\delta h_{bd}+\frac{1}{16\pi G}\int_{\mathcal{S}}d^{3}x\sqrt{-h}n_{a}\nabla_{c}(\theta_{CS}^{~~~*}R^{abcd})\delta h_{bd}\\
\nonumber&\simeq&-\frac{1}{16\pi G}\int_{\mathcal{S}}d^{3}x\sqrt{-h}\nabla_{a}(n_{c}\theta_{CS}^{~~~*}R^{abcd}\delta h_{bd})+\frac{1}{16\pi G}\int_{\mathcal{S}}d^{3}x\sqrt{-h}\nabla_{a}(n_{c}\theta_{CS}^{~~~*}R^{abcd})\delta h_{bd}\\
\nonumber&&+\frac{1}{16\pi G}\int_{\mathcal{S}}d^{3}x\sqrt{-h}n_{a}\nabla_{c}(\theta_{CS}^{~~~*}R^{abcd})\delta h_{bd}\\
\nonumber&\simeq&\frac{1}{16\pi G}\int_{\mathcal{S}}d^{3}x\sqrt{-h}\nabla_{a}(n_{c}\theta_{CS}^{~~~*}R^{abcd})\delta h_{bd}+\frac{1}{16\pi G}\int_{\mathcal{S}}d^{3}x\sqrt{-h}n_{a}\nabla_{c}(\theta_{CS}^{~~~*}R^{abcd})\delta h_{bd}\\
\nonumber&\simeq&-\int_{\mathcal{S}}d^{3}x\sqrt{-h}\bar{t}_{CS}^{b'd'}(h_{b'}^{~b}+n_{b'}n^{b})(h_{d'}^{~d}+n_{d'}n^{d})\delta h_{bd}\\
&\simeq&-\int_{\mathcal{S}}d^{3}x\sqrt{-h}\bar{t}_{CS}^{b'd'}h_{b'}^{~b}h_{d'}^{~d}\delta h_{bd}.
\end{eqnarray}
The first term in the second line vanishes for the following reason. According to the Gauss theorem, this term can be put into a 2-dimensional integral on the boundary $\partial\mathcal{S}$ of stretched horizon. Since the stretched horizon has no boundary, this term vanishes. For brevity, we have introduced the notation
\begin{equation}\label{tbarcs}
\bar{t}_{CS}^{bd}=-\frac{1}{32\pi G} \epsilon^{cdef}[\nabla_{a}(n_{c}\theta_{CS} R^{ab}_{~~ef})+n_{a}\nabla_{c}(\theta_{CS} R^{ab}_{~~ef})]+(b\leftrightarrow d),
\end{equation}
where $\bar{t}^{bd}$ is symmetric with respect to indices $b$ and $d$. In the last line of equation \eqref{vScs}, we have made use of $n^{d}h_{bd}=0$ and $n^{d}\delta h_{bd}=-h_{bd}\delta n^{d}=0$ on the stretched horizon.

Comparing the above expression of $\delta S_{CS}$ with
\begin{equation}
\delta S_{CS}=-\frac{1}{2}\int_{\mathcal{S}}d^{3}x\sqrt{h} t_{CS}^{bd}\delta h_{bd},
\end{equation}
we find\footnote{Naively, one might have taken $\bar{t}_{CS}^{bd}$ as the membrane stress tensor. But it is provable that $\bar{t}_{CS}^{bd}n_{d}\neq0$, thus $\bar{t}_{CS}^{bd}$ cannot be decomposed as equation \eqref{compose}.}
\begin{equation}\label{tcs}
t_{CS}^{bd}=h^{b}_{~b'}h^{d}_{~d'}\bar{t}_{CS}^{b'd'}.
\end{equation}
Combined with equation \eqref{teh}, it gives the total stress tensor of the $(2+1)$-dimensional fluid on the stretched horizon
\begin{eqnarray}\label{ttot}
\nonumber t^{bd}&=&t_{EH}^{bd}+t_{CS}^{bd}\\
\nonumber &=&\frac{1}{8\pi G}(Kh^{bd}-K^{bd})-\frac{1}{32\pi G}h^{b}_{~b'}h^{d}_{~d'}\epsilon^{cd'ef}[\nabla_{a}(n_{c}\theta_{CS} R^{ab'}_{~~ef})+n_{a}\nabla_{c}(\theta_{CS} R^{ab'}_{~~ef})]\\
&&-\frac{1}{32\pi G}h^{b}_{~b'}h^{d}_{~d'}\epsilon^{cb'ef}[\nabla_{a}(n_{c}\theta_{CS} R^{ad'}_{~~ef})+n_{a}\nabla_{c}(\theta_{CS} R^{ad'}_{~~ef})].
\end{eqnarray}

\subsection{Membrane fluid equations}\label{subsect-flueq}
From the Gauss-Codazzi equations \cite{Gourgoulhon:2005ng}, it is possible to prove that the membrane stress tensor satisfies
\begin{equation}\label{Gauss-Codazzi}
8\pi Gt^{bd}_{EH|d}=-h^{b}_{~c}R^{cd}n_{d},
\end{equation}
where $|d$ is the 3-covariant derivative with respect to the metric $h_{bd}$. Together with the Einstein equation $R^{bd}-g^{bd}R/2=8\pi GT^{bd}$, it leads to a relation between the membrane stress tensor and bulk stress tensor \cite{Parikh:1997ma},
\begin{equation}\label{mbrel}
t^{bd}_{~~|d}=-h^{b}_{~c}T^{cd}n_{d}.
\end{equation}
This relation will be dubbed membrane-bulk relation in our paper, and it plays a central role in deriving the fluid equations in the membrane paradigm. Unfortunately, it is very difficult to prove a membrane-bulk relation of the same form in Chern-Simons modified gravity generally, but this can be done for some specific background geometries, e.g., the spherical static spacetime in the next section.

On the other hand, because $t^{bd}n_{d}=0$, we can decompose the above stress tensor to a form like a viscous fluid with a Hall viscosity term \cite{Saremi:2011ab,Jensen:2011xb},
\begin{equation}\label{compose}
t^{bd}=\frac{1}{\alpha}\rho u^{b}u^{d}+\frac{1}{\alpha}\gamma^{b}_{~A}\gamma^{d}_{~B}(p\gamma^{AB}-2\eta\sigma^{AB}-\zeta\theta\gamma^{AB}-2\tilde{\eta}\tilde{\sigma}^{AB})+\pi^{A}(\gamma^{~b}_{A}u^{d}+\gamma^{~d}_{A}u^{b}).
\end{equation}
Here we have inserted the renormalization parameter $\alpha$, hence $\rho$, $p$, $\sigma^{AB}$, $\theta$, $\tilde{\sigma}^{AB}$ and $\pi^{A}$ correspond to fluid quantities on the event horizon, although $t^{bd}$ is the fluid stress tensor on the stretched horizon.

Combining \eqref{mbrel} and \eqref{compose}, and taking the limit $\alpha\rightarrow 0$, we will get the $(2+1)$-dimensional fluid continuity equation and the Navier-Stokes equation \cite{Wang:2014aty,Gourgoulhon:2005ng,Gourgoulhon:2005ch},
\begin{eqnarray}
\mathcal{L}_{l}\rho+\theta\rho&=&-p\theta+2\eta\sigma_{AB}\sigma^{AB}+\zeta\theta^{2}+2\tilde{\eta}\tilde{\sigma}_{AB}\tilde{\sigma}^{AB}+T^{a}_{~b}l_{a}l^{b},\label{Ray}\\
\gamma_{A}^{~e}\mathcal{L}_{l}\pi_{e}+\pi_{A}\theta&=&-p_{||A}+2(\eta\sigma^{B}_{~A})_{||B}+(\zeta\theta)_{||A}+2(\tilde{\eta}\tilde{\sigma}^{B}_{~A})_{||B}-T^{c}_{~a}l_{c}\gamma^{a}_{~A},\label{NS}
\end{eqnarray}
where $||A$ is the 2-covariant derivative with respect to the metric $\gamma_{AB}$.

\section{Application to spherical static spacetime}\label{sect-sphst}
Without loss of generality, we can use two functions $f_{1}(r)$ and $f_{2}(r)$ to write down the metric of spherically symmetric static spacetime
\begin{equation}\label{metric}
ds^{2}=-f_{1}(r)dt^{2}+f_{2}(r)dr^{2}+r^{2}(d\vartheta^{2}+\sin^{2}\vartheta d\varphi^{2}).
\end{equation}
If there is a membrane paradigm for this spacetime, the spacelike normal vector and the timelike generator of the stretched horizon will take the form
\begin{equation}\label{nu}
n^{a}\partial_{a}=f_{2}^{-1/2}\partial_{r},~~~~u^{a}\partial_{a}=f_{1}^{-1/2}\partial_{t}
\end{equation}
One may readily check that the H\'a\'{\j}i\v{c}ek field vanishes $\Omega_{A}=0$ in this situation. According to the definition of the 3-dimensional extrinsic curvature, we get the nonvanishing components
\begin{equation}\label{}
K_{tt}=-\frac{f_{1}'}{2\sqrt{f_{2}}},~~~~K_{\vartheta\vartheta}=\frac{r}{\sqrt{f_{2}}},~~~~K_{\varphi\varphi}=\frac{r\sin^{2}\vartheta}{\sqrt{f_{2}}},
\end{equation}
in which $'$ indicates the derivative with respect to $r$. We can also find that
\begin{eqnarray}
\nonumber&&\gamma^{a}_{~A}\gamma^{b}_{~B}\nabla_{b}u_{a}=\gamma^{a}_{~A}n^{b}\nabla_{b}u_{a}=\gamma^{b}_{~B}n^{a}\nabla_{b}u_{a}=0,\\
\nonumber&&\gamma^{a}_{~A}u^{b}\nabla_{b}u_{a}=\gamma^{b}_{~B}u^{a}\nabla_{b}u_{a}=n^{b}u^{a}\nabla_{b}u_{a}=0,\\
&&u^{b}n^{a}\nabla_{b}u_{a}=\frac{f_{1}'}{2f_{1}\sqrt{f_{2}}}.
\end{eqnarray}
Starting from equation \eqref{metric}, we can get all terms of the Riemann tensor as below,
\begin{eqnarray}
\nonumber &&R^{tr}_{~~tr}=-R^{rt}_{~~tr}=-R^{tr}_{~~rt}=R^{rt}_{~~rt}=\frac{f_{1}f_{1}'f_{2}'+f_{2}(f'^{2}_{1}-2f_{1}f_{1}'')}{4f_{1}^{2}f_{2}^{2}}, \\
\nonumber &&R^{t\vartheta}_{~~t\vartheta}=-R^{\vartheta t}_{~~t\vartheta}=-R^{t\vartheta}_{~~\vartheta t}=R^{\vartheta t}_{~~\vartheta t}=R^{t\varphi}_{~~t\varphi}=-R^{\varphi t}_{~~t\varphi}=-R^{t\varphi}_{~~\varphi t}=R^{\varphi t}_{~~\varphi t}=-\frac{f_{1}'}{2rf_{1}f_{2}}, \\
\nonumber &&R^{r\vartheta}_{~~r\vartheta}=-R^{\vartheta r}_{~~r\vartheta}=-R^{r\vartheta}_{~~\vartheta r}=R^{\vartheta r}_{~~\vartheta r}=R^{r\varphi}_{~~r\varphi}=-R^{\varphi r}_{~~r\varphi}=-R^{r\varphi}_{~~\varphi r}=R^{\varphi r}_{~~\varphi r}=\frac{f_{2}'}{2rf_{2}^{2}}, \\
&&R^{\vartheta\varphi}_{~~\vartheta\varphi}=-R^{\varphi\vartheta}_{~~\vartheta\varphi}=-R^{\vartheta\varphi}_{~~\varphi\vartheta}=R^{\varphi\vartheta}_{~~\varphi\vartheta}=\frac{f_{2}-1}{r^{2}f_{2}}.
\end{eqnarray}

These results seem to be formidably clumsy for proceeding to build the membrane paradigm, including fluid quantities and fluid equations. Fortunately, this can be accomplished at the background level and has the potential to be extended perturbatively\footnote{Actually, by studying linear perturbations of $\tilde{\sigma}^{AB}$, we have obtained the Hall viscosity $\tilde{\eta}$ in subsections \ref{subsect-fluqn} and \ref{subsect-sch}.}. The key observation is that the above expressions of $K_{ab}$, $\nabla_{a}u_{b}$ and $R^{ab}_{~~ef}$ appear in the elegant form
\begin{eqnarray}\label{KR}
\nonumber K_{ab}&=&\lambda_1(r)u_{a}u_{b}+\lambda_2(r)n_{a}n_{b}+\lambda_3(r)g_{ab},\\
\nonumber \nabla_{a}u_{b}&=&\lambda_4(r)u_{a}n_{b},\\
\nonumber R^{ab}_{~~ef}&=&\lambda_5(r)(u^{a}u_{e}n^{b}n_{f}-u^{b}u_{e}n^{a}n_{f}-u^{a}u_{f}n^{b}n_{e}+u^{b}u_{f}n^{a}n_{e})\\
\nonumber&&+\lambda_6(r)(u^{a}u_{e}g^{b}_{~f}-u^{b}u_{e}g^{a}_{~f}-u^{a}u_{f}g^{b}_{~e}+u^{b}u_{f}g^{a}_{~e})\\
\nonumber&&+\lambda_7(r)(n^{a}n_{e}g^{b}_{~f}-n^{b}n_{e}g^{a}_{~f}-n^{a}n_{f}g^{b}_{~e}+n^{b}n_{f}g^{a}_{~e})\\
&&+\lambda_8(r)(g^{a}_{~e}g^{b}_{~f}-g^{b}_{~e}g^{a}_{~f}),
\end{eqnarray}
in which
\begin{eqnarray}
\nonumber&&\lambda_{1}=\frac{2f_{1}-rf^{'}_{1}}{2rf_{1}\sqrt{f_{2}}},~~~~\lambda_{2}=-\frac{1}{r\sqrt{f_{2}}},~~~~\lambda_{3}=\frac{1}{r\sqrt{f_{2}}},~~~~\lambda_{4}=-\frac{f^{'}_{1}}{2f_{1}\sqrt{f_{2}}},\\
\nonumber&&\lambda_5=-\frac{f_{1}f_{1}'f_{2}'+f_{2}(f'^{2}_{1}-2f_{1}f_{1}'')}{4f_{1}^{2}f_{2}^{2}}-\frac{f_{1}'}{2rf_{1}f_{2}}+\frac{f_{2}'}{2rf_{2}^{2}}-\frac{f_{2}-1}{r^{2}f_{2}},\\
&&\lambda_6=\frac{f_{1}'}{2rf_{1}f_{2}}+\frac{f_{2}-1}{r^{2}f_{2}},~~~~\lambda_7=\frac{f_{2}'}{2rf_{2}^{2}}-\frac{f_{2}-1}{r^{2}f_{2}},~~~~\lambda_8=\frac{f_{2}-1}{r^{2}f_{2}}.
\end{eqnarray}
It is worthwhile to note that $\lambda_{2}+\lambda_{3}=0$, which is in accordance with the normal condition $n^{a}K_{ab}=0$.

Now it is possible to compute the stress tensor \eqref{ttot}. In preparation, we work out the following equalities:
\begin{eqnarray}
\nonumber\epsilon^{cdef}R^{ab}_{~~ef}&=&2\lambda_5\epsilon^{cdef}(u^{a}u_{e}n^{b}n_{f}-u^{b}u_{e}n^{a}n_{f})+2\lambda_6\epsilon^{cdef}(u^{a}u_{e}g^{b}_{~f}-u^{b}u_{e}g^{a}_{~f})\\
\label{epsR}&&+2\lambda_7\epsilon^{cdef}(n^{a}n_{e}g^{b}_{~f}-n^{b}n_{e}g^{a}_{~f})+2\lambda_8\epsilon^{cdef}g^{a}_{~e}g^{b}_{~f},\\
n_{c}\epsilon^{cdef}R^{ab}_{~~ef}&=&2\lambda_6n_{c}\epsilon^{cdeb}u^{a}u_{e}-2\lambda_6n_{c}\epsilon^{cdea}u^{b}u_{e}+2\lambda_8n_{c}\epsilon^{cdab},\\
n_{a}\epsilon^{cdef}R^{ab}_{~~ef}&=&-2(\lambda_5+\lambda_6)\epsilon^{cdef}u^{b}u_{e}n_{f}+2(\lambda_7+\lambda_8)\epsilon^{cdeb}n_{e},\\
K_{ca}\epsilon^{cdef}R^{ab}_{~~ef}&=&2\lambda_5K_{ca}\epsilon^{cdef}u^{a}u_{e}n^{b}n_{f}+2\lambda_6K_{ca}\epsilon^{cdeb}u^{a}u_{e}.
\end{eqnarray}
Substituting them into equations \eqref{tbarcs} and \eqref{tcs}, we can get
\begin{eqnarray}
\nonumber8\pi G\bar{t}_{CS}^{bd}&=&-\frac{1}{4}\left[\nabla_{a}(n_{c}\theta_{CS}\epsilon^{cdef}R^{ab}_{~~ef})+\nabla_{c}(n_{a}\theta_{CS}\epsilon^{cdef}R^{ab}_{~~ef})-K_{ca}\theta_{CS}\epsilon^{cdef}R^{ab}_{~~ef}\right]+(b\leftrightarrow d)\\
\nonumber&=&\frac{1}{2}\bigl\{\nabla_{a}(\lambda_6n_{c}\theta_{CS}\epsilon^{cdea}u^{b}u_{e})+\nabla_{a}[(\lambda_5+\lambda_6)\theta_{CS}\epsilon^{adec}u^{b}u_{e}n_{c}]\\
\nonumber&&+\lambda_5(\lambda_1u_{c}u_{a}+\lambda_2n_{c}n_{a}+\lambda_3g_{ca})\theta_{CS}\epsilon^{cdef}u^{a}u_{e}n^{b}n_{f}\bigr\}+(b\leftrightarrow d)\\
\nonumber&=&\frac{1}{2}\nabla_{a}(\lambda_5\theta_{CS}\epsilon^{adec}u^{b}u_{e}n_{c})+(b\leftrightarrow d)\\
&=&\frac{1}{2}\nabla_{a}(\lambda_5\theta_{CS})\epsilon^{adec}u^{b}u_{e}n_{c}+(b\leftrightarrow d)
\end{eqnarray}
and consequently
\begin{equation}\label{tcssphst}
8\pi Gt_{CS}^{bd}=\frac{1}{2}\nabla_{a}(\lambda_5\theta_{CS})\epsilon^{adec}u^{b}u_{e}n_{c}+(b\leftrightarrow d).
\end{equation}
Moreover, remembering that $\lambda_{2}+\lambda_{3}=0$, we can obtain the Einstein-Hilbert part of the stress tensor
\begin{equation}\label{}
t_{EH}^{bd}=\frac{1}{8\pi G}(2\lambda_{3}-\lambda_{1})\gamma^{bd}-\frac{1}{4\pi G}\lambda_{3}u^{b}u^{d}.
\end{equation}
As a result, the full expression of the membrane stress tensor is
\begin{eqnarray}\label{tsphst}
t^{bd}=\frac{1}{8\pi G}(2\lambda_{3}-\lambda_{1})\gamma^{bd}-\frac{1}{4\pi G}\lambda_{3}u^{b}u^{d}+\frac{1}{16\pi G}\nabla_{a}(\lambda_5\theta_{CS})(\epsilon^{adec}u^{b}+\epsilon^{abec}u^{d})u_{e}n_{c}.
\end{eqnarray}
This result is much more tamable than equation \eqref{ttot}. With it in hand, we can proceed to put the membrane paradigm in a concrete form.

\subsection{Membrane-bulk relation}\label{subsect-mbrel}
The membrane-bulk relation \eqref{mbrel} is crucial for the derivation of membrane fluid equations. In this subsection, we will prove that the stress tensor \eqref{tsphst} can satisfy this equation.

In the Chern-Simons modified gravity, the gravitational field equation can be written as
\begin{equation}\label{}
R^{ab}-\frac{1}{2}g^{ab}R+C^{ab}=8 \pi G T^{ab}.
\end{equation}
The symmetric, traceless Cotton tensor $C^{ab}$ is defined by \cite{Grumiller:2008ie}
\begin{equation}\label{cotton}
C^{ab}=\frac{1}{2}[(\nabla_{c}\theta_{CS})\epsilon^{cdea}\nabla_{e}R^{b}_{~d}+(\nabla_{d}\nabla_{c}\theta_{CS})^{*}R^{cabd}]+(a\leftrightarrow b).
\end{equation}
Keeping in mind that $n_{b}h^{ba}=0$, to satisfy the membrane-bulk relation \eqref{mbrel}, we just need to prove
\begin{equation}\label{con}
8\pi Gt^{ab}_{~~\mid b}=-(R^{cb}+C^{cb})n_{b}h_{c}^{~a}.
\end{equation}

From equation \eqref{tcssphst}, we can get
\begin{eqnarray}
\nonumber8\pi Gt^{ij}_{CS|j}&=&\frac{1}{2}h^{i}_{~b}h^{j}_{~d}\nabla_{j}[\nabla_{a}(\lambda_5\theta_{CS})\epsilon^{adec}u^{b}u_{e}n_{c}+\nabla_{a}(\lambda_5\theta_{CS})\epsilon^{abec}u^{d}u_{e}n_{c}]\\
\nonumber&=&\frac{1}{2}h^{i}_{~b}h^{j}_{~d}\nabla_{j}[\nabla_{a}(\lambda_5\theta_{CS})\epsilon^{adec}u^{b}u_{e}n_{c}+\nabla_{a}(\lambda_5\theta_{CS})\epsilon^{abec}u^{d}u_{e}n_{c}]\\
\nonumber&=&\frac{1}{2}h^{i}_{~b}h^{j}_{~d}\epsilon^{adec}\bigl[u^{b}u_{e}n_{c}\nabla_{j}\nabla_{a}(\lambda_{5}\theta_{CS})+u_{e}n_{c}\nabla_{a}(\lambda_{5}\theta_{CS})\nabla_{j}u^{b}\\
\nonumber&&+u^{b}n_{c}\nabla_{a}(\lambda_{5}\theta_{CS})\nabla_{j}u_{e}+u_{e}u^{b}\nabla_{a}(\lambda_{5}\theta_{CS})\nabla_{j}n_{c}\bigr]\\
\nonumber&&+\frac{1}{2}h^{i}_{~b}h^{j}_{~d}\epsilon^{abec}\bigl[u^{d}u_{e}n_{c}\nabla_{j}\nabla_{a}(\lambda_{5}\theta_{CS})+u_{e}n_{c}\nabla_{a}(\lambda_{5}\theta_{CS})\nabla_{j}u^{d}\\
\nonumber&&+u^{d}n_{c}\nabla_{a}(\lambda_{5}\theta_{CS})\nabla_{j}u_{e}+u_{e}u^{d}\nabla_{a}(\lambda_{5}\theta_{CS})\nabla_{j}n_{c}\bigr]\\
&=&\frac{1}{2}\epsilon^{ajec}u^{i}u_{e}n_{c}\nabla_{j}\nabla_{a}(\lambda_{5}\theta_{CS})+\frac{1}{2}\epsilon^{aiec}u^{j}u_{e}n_{c}\nabla_{j}\nabla_{a}(\lambda_{5}\theta_{CS}).
\end{eqnarray}
In the last step, we have employed equations \eqref{KR} and the antisymmetry of $\epsilon^{abcd}$. Note that $\nabla_{j}\nabla_{a}(\lambda_{5}\theta_{CS})$ is symmetric with respect to indices $j$ and $a$, hence we can further get
\begin{equation}\label{spt}
8\pi Gt^{ab}_{CS\mid b}=\frac{1}{2}\epsilon^{daec}u^{b}u_{e}n_{c}\nabla_{b}\nabla_{d}(\lambda_{5}\theta_{CS}).
\end{equation}

As shown in appendix \ref{app-Cnh}, it is possible to demonstrate that
\begin{equation}\label{spc}
C^{cb}n_{b}h_{c}^{~a}=-\frac{1}{2}\epsilon^{cade}u^{f}u_{d}n_{e}\nabla_{f}\nabla_{c}(\lambda_{5}\theta_{CS}).
\end{equation}
From \eqref{spt} with \eqref{spc}, it is trivial to read
\begin{equation}\label{}
8\pi Gt^{ab}_{CS\mid b}=-C^{cb}n_{b}h_{c}^{~a}.
\end{equation}
Combined with equation \eqref{Gauss-Codazzi}, this completes the proof of equation \eqref{con} and hence the membrane-bulk relation.

\subsection{Fluid quantities}\label{subsect-fluqn}
From equation \eqref{tsphst}, one can check that
\begin{eqnarray}\label{}
\nonumber &&u_{b}u_{d}t^{bd}=-\frac{1}{4\pi G}\lambda_{3},~~~~\gamma^{A}_{~b}\gamma^{B}_{~d}t^{bd}=\frac{1}{8\pi G}(2\lambda_{3}-\lambda_{1})\gamma^{AB},\\
&&\gamma^{A}_{~b}u_{d}t^{bd}=-\frac{1}{16\pi G}\gamma^{A}_{~b}\nabla_{a}(\lambda_5\theta_{CS})\epsilon^{abec}u_{e}n_{c}.
\end{eqnarray}

According to equation \eqref{compose}, we thus have
\begin{eqnarray}\label{}
\nonumber&&\rho=-\frac{\alpha}{4\pi G}\lambda_{3},~~~~\pi^{A}=\frac{1}{16\pi G}\gamma^{A}_{~b}\nabla_{a}(\lambda_5\theta_{CS})\epsilon^{abec}u_{e}n_{c},\\
&&p\gamma^{AB}-2\eta\sigma^{AB}-\zeta\theta\gamma^{AB}-2\tilde{\eta}\tilde{\sigma}^{AB}=\frac{\alpha}{8\pi G}(2\lambda_{3}-\lambda_{1})\gamma^{AB}.
\end{eqnarray}
In order to fulfill the limit \eqref{Klim}, at the background level we should impose the following conditions \cite{Wang:2014aty}:
\begin{equation}\label{}
\eta=\frac{1}{16\pi G},~~~~\theta=2\alpha\lambda_{3},~~~~\zeta=-\frac{1}{16\pi G}.
\end{equation}
The appearance of antisymmetric tensor $\epsilon^{abcd}$ in $t_{CS}^{bd}$ is remarkable. It implies that the Chern-Simons correction does not change the shear viscosity $\eta$ but induces the Hall viscosity $\tilde{\eta}$.

As a result, the fluid quantities on the event horizon are
\begin{eqnarray}\label{fluqn}
\nonumber&&\mathrm{Energy~density:~}\rho=-\frac{\alpha}{4\pi G}\lambda_{3},\\
\nonumber&&\mathrm{Pressure:~}p=\frac{\alpha}{8\pi G}(\lambda_{3}-\lambda_{1}),\\
\nonumber&&\mathrm{Momentum~density:~}\pi^{A}=\frac{1}{16\pi G}\gamma^{A}_{~b}\nabla_{a}(\lambda_5\theta_{CS})\epsilon^{abef}u_{e}n_{f},\\
\nonumber&&\mathrm{Shear:~}\sigma^{AB}=0,\\
\nonumber&&\mathrm{Expansion:~}\theta=2\alpha\lambda_{3},\\
\nonumber&&\mathrm{Shear~viscosity:~}\eta=\frac{1}{16\pi G},\\
&&\mathrm{Bulk~viscosity:~}\zeta=-\frac{1}{16\pi G}.
\end{eqnarray}

Because $\sigma^{AB}=0$, by definition \eqref{sigtd} we get $\tilde{\sigma}^{AB}=0$. It is then impossible to extract $\tilde{\eta}$ from the projection $\tilde{\sigma}_{bd}t^{bd}=-2\alpha^{-1}\tilde{\eta}\tilde{\sigma}_{AB}\tilde{\sigma}^{AB}$. Instead, by brute force we expand the stress tensor \eqref{ttot} in the limit \eqref{Kdulim}, and pick out terms proportional to $\tilde{\sigma}^{bd}$ and hence $\tilde{\eta}$ from the coefficient. The calculation is tedious. Here we write down the final result.
\begin{eqnarray}\label{Hav1}
\nonumber\mathrm{Hall~viscosity:~}\tilde{\eta}&=&\frac{1}{16\pi G}\biggl\{\theta_{CS}\mathcal{K}+\alpha u^{a}\nabla_{a}\left[\theta_{CS}\alpha^{-2}\left(\theta-g_{\mathcal{H}}\right)\right]+2\alpha n^{a}\nabla_{a}\left(\theta_{CS}\alpha^{-2}g_{\mathcal{H}}\right)\\
&&+\theta_{CS}\alpha^{-2}\left(\frac{1}{4}\theta^2+g_{\mathcal{H}}\theta+2g_{\mathcal{H}}^2\right)+\alpha u^{a}\nabla_{a}\left[\theta_{CS}n^{c}\nabla_{c}\left(\alpha^{-1}\right)\right]\biggr\}.
\end{eqnarray}
In the above, $g_{\mathcal{H}}$ is the surface gravity at the event horizon. $\mathcal{K}$ is the Gaussian curvature  of the 2-dimensional spacelike section of the event horizon. In terms of $\mathcal{K}$, the 2-dimensional Riemann tensor can be expressed as ${^{(2)}}R^{ab}_{~~ef}=\mathcal{K}\left(\gamma^{a}_{~e}\gamma^{b}_{~f}-\gamma^{a}_{~f}\gamma^{b}_{~e}\right)$. A careful comparison of our result with \cite{Saremi:2011ab} and the details of calculation will be reported in a successive article \cite{wang:2016}.

It is interesting to compare this result with the membrane paradigm in Einstein gravity. Besides the Hall viscosity, they are also different in the momentum density. In the Einstein theory, the momentum density is zero on the stretched horizon in spherically symmetric spacetimes. In the Chern-Simons modified gravity, the momentum density could be nonvanishing if the Chern-Simons scalar $\theta_{CS}$ is nonconstant. In fact, under the convention $\epsilon_{tr\vartheta\varphi}=\sqrt{-g}$, the nonzero components are
\begin{equation}\label{}
\pi^{\vartheta}=-\frac{1}{16\pi Gr^2\sin\vartheta}\partial_{\varphi}(\lambda_{5}\theta_{CS}),~~~~\pi^{\varphi}=\frac{1}{16\pi Gr^2\sin\vartheta}\partial_{\vartheta}(\lambda_{5}\theta_{CS}).
\end{equation}
In the next subsection, we will consider a specific example with $\pi^{A}\neq0$.

\subsection{Example: Schwarzschild black hole}\label{subsect-sch}
In reference \cite{Grumiller:2007rv}, it has been verified that the Schwarzschild line element
\begin{equation}\label{metricsch}
ds^{2}=-\left(1-\frac{2GM}{r}\right)dt^{2}+\left(1-\frac{2GM}{r}\right)^{-1}dr^{2}+r^{2}(d\vartheta^{2}+\sin^{2}\vartheta d\varphi^{2})
\end{equation}
is always a solution of the Chern-Simons modified gravity if the scalar is of the form
\begin{equation}\label{FrG}
\theta_{CS}=\mathcal{F}(t,r)+r\mathcal{G}(\vartheta,\varphi).
\end{equation}

Accordingly the spacelike normal vector and the timelike generator of the stretched horizon take the form
\begin{equation}\label{}
n^{a}\partial_{a}=\left(1-\frac{2GM}{r}\right)^{1/2}\partial_{r},~~~~u^{a}\partial_{a}=\left(1-\frac{2GM}{r}\right)^{-1/2}\partial_{t},
\end{equation}
the renormalization parameter becomes
\begin{equation}\label{}
\alpha=\left(1-\frac{2GM}{r}\right)^{1/2},
\end{equation}
and the nonvanishing components of $K_{ab}$, $\nabla_{a}u_{b}$ and $R^{ab}_{~~ef}$  are given by
\begin{eqnarray}
\nonumber&&K_{tt}=-\frac{GM}{r^{2}}\left(1-\frac{2GM}{r}\right)^{1/2},~~~~ K_{\vartheta\vartheta}=r\left(1-\frac{2GM}{r}\right)^{1/2},\\
\nonumber&&K_{\varphi\varphi}=r\sin^{2}\vartheta\left(1-\frac{2GM}{r}\right)^{1/2},~~~~u^{b}n^{a}\nabla_{b}u_{a}=\frac{GM}{r^{2}}\left(1-\frac{2GM}{r}\right)^{-1/2},\\
\nonumber&&R^{tr}_{~~tr}=-R^{rt}_{~~tr}=-R^{tr}_{~~rt}=R^{rt}_{~~rt}=\frac{2GM}{r^{3}},\\
\nonumber&&R^{t\vartheta}_{~~t\vartheta}=-R^{\vartheta t}_{~~t\vartheta}=-R^{t\vartheta}_{~~\vartheta t}=R^{\vartheta t}_{~~\vartheta t}=R^{t\varphi}_{~~t\varphi}=-R^{\varphi t}_{~~t\varphi}=-R^{t\varphi}_{~~\varphi t}=R^{\varphi t}_{~~\varphi t}=-\frac{GM}{r^{3}},\\
\nonumber &&R^{r\vartheta}_{~~r\vartheta}=-R^{\vartheta r}_{~~r\vartheta}=-R^{r\vartheta}_{~~\vartheta r}=R^{\vartheta r}_{~~\vartheta r}=R^{r\varphi}_{~~r\varphi}=-R^{\varphi r}_{~~r\varphi}=-R^{r\varphi}_{~~\varphi r}=R^{\varphi r}_{~~\varphi r}=-\frac{GM}{r^{3}},\\
&&R^{\vartheta\varphi}_{~~\vartheta\varphi}=-R^{\varphi\vartheta}_{~~\vartheta\varphi}=-R^{\vartheta\varphi}_{~~\varphi\vartheta}=R^{\varphi\vartheta}_{~~\varphi\vartheta}=\frac{2GM}{r^{3}}.
\end{eqnarray}
In other words, $\lambda_{i}$'s in equations \eqref{KR} are
\begin{eqnarray}
\nonumber&&\lambda_{1}=\frac{r-3GM}{r^{2}}\left(1-\frac{2GM}{r}\right)^{-1/2},~~~~\lambda_{2}=-\frac{1}{r}\left(1-\frac{2GM}{r}\right)^{1/2},\\
\nonumber&&\lambda_{3}=\frac{1}{r}\left(1-\frac{2GM}{r}\right)^{1/2},~~~~\lambda_{4}=-\frac{GM}{r^2}\left(1-\frac{2GM}{r}\right)^{-1/2},\\
&&\lambda_5=-\frac{6GM}{r^{3}},~~~~\lambda_6=\frac{3GM}{r^{3}},~~~~\lambda_7=-\frac{3GM}{r^{3}},~~~~\lambda_8=\frac{2GM}{r^{3}}.
\end{eqnarray}

For the choice of $\theta_{CS}$ given in equation \eqref{FrG}, we can get the membrane stress tensor
\begin{eqnarray}\label{}
\nonumber t^{bd}&=&\frac{r-GM}{8\pi Gr^{2}}\left(1-\frac{2GM}{r}\right)^{-1/2}\gamma^{bd}-\frac{1}{4\pi Gr}\left(1-\frac{2GM}{r}\right)^{1/2}u^{b}u^{d}\\
&&-\frac{1}{16\pi G}\nabla_{a}\left[\frac{6GM(\mathcal{F}+r\mathcal{G})}{r^{3}}\right](\epsilon^{adec}u^{b}+\epsilon^{abec}u^{d})u_{e}n_{c}
\end{eqnarray}
as well as the expansion, energy density, fluid pressure and momentum density
\begin{eqnarray}
\nonumber&&\theta=\frac{2}{r}\left(1-\frac{2GM}{r}\right),~~~~\rho=-\frac{1}{4\pi Gr}\left(1-\frac{2GM}{r}\right),~~~~p=\frac{M}{8\pi r^{2}},\\
&&\pi^{\vartheta}=\frac{3M}{8\pi r^4\sin\vartheta}\partial_{\varphi}\mathcal{G},~~~~\pi^{\varphi}=-\frac{3M}{8\pi r^4\sin\vartheta}\partial_{\vartheta}\mathcal{G},
\end{eqnarray}
while other quantities are zero or nonzero constants given by \eqref{fluqn}. As has done in reference \cite{Wang:2014aty}, one can take $l^{a}\partial_{a}=(\partial_{t}+f\partial_{r})/2$ and check that fluid equations \eqref{Ray} and \eqref{NS} are satisfied to the leading order in the limit $\alpha\rightarrow0$, even though the momentum density $\pi^{A}$ is nonvanishing now. The nonvanishness of $\pi^{A}$ is counterintuitive, because the H\'a\'{\j}i\v{c}ek field $\Omega_{A}=0$ for Schwarzschild metric. A full investigation of this behavior in the dynamical Chern-Simons gravity is desired. To avoid distraction, we leave it for future investigation.

In the present example, the Hall viscosity can be rearranged as
\begin{eqnarray}\label{Hav2}
\nonumber\tilde{\eta}&=&\frac{1}{16\pi G}\biggl\{\theta_{CS}\mathcal{K}+\partial_{t}\left[\theta_{CS}\alpha^{-2}\left(\theta-g_{\mathcal{H}}\right)\right]+2\alpha^2\partial_{r}\left(\theta_{CS}\alpha^{-2}g_{\mathcal{H}}\right)\\
&&+\theta_{CS}\alpha^{-2}\left(\frac{1}{4}\theta^2+g_{\mathcal{H}}\theta+2g_{\mathcal{H}}^2\right)-\frac{1}{2}\partial_{t}\left[\theta_{CS}\alpha^{-2}\partial_{r}\left(\alpha^2\right)\right]\biggr\},
\end{eqnarray}
in which $\mathcal{K}=1/r^2$ and $g_{\mathcal{H}}=GM/r^2$. This expression indicates that the Hall viscosity is divergent unless $\theta_{CS}\sim\mathcal{O}(\alpha^2)$ as $\alpha\rightarrow0$. For this reason, we take condition \eqref{FrG} into account and make the ansatz $\theta_{CS}=F(t,r)\alpha^2$, where $F(t,r)$ is finite in the limit $\alpha\rightarrow0$. Under this ansatz, the leading order contributions to Hall viscosity are
\begin{equation}
\tilde{\eta}=\frac{1}{16\pi G}\left(-\frac{2GM}{r^2}\partial_{t}F+\frac{2G^2M^2F}{r^4}\right).
\end{equation}
As a concrete example, if we set $F=r^2$, namely $\theta_{CS}=r^2-2GMr$, this will yield $\tilde{\eta}=1/(32\pi G)$. In reference \cite{Saremi:2011ab}, the Hall viscosity has been evaluated in Chern-Simons gravity in a different method for a different background spacetime\footnote{Their assumption (3.17) excludes our metric \eqref{metricsch}.}. Their result corresponds to our third term of \eqref{Hav2}, which is apparently negligible compared with other terms. A careful comparison will be made in a successive paper \cite{wang:2016}.

\section{Outlook}\label{sect-outl}
In this paper, we have studied the membrane paradigm for black holes in the Chern-Simons modified gravity. We tried our best to keep up the details. For instance, relation \eqref{mbrel} is crucial for the derivation of membrane fluid equations, so we scrutinized this relation carefully in the Chern-Simons gravity theory. Clearly, the membrane paradigm can be established reliably in this theory, at least for the spherical static case.

To our surprise, although the H\'a\'{\j}i\v{c}ek field vanishes in the spherically static spacetime, the fluid momentum density on the membrane is nonzero if the dynamical field $\theta_{CS}$ varies with angular coordinates $(\vartheta,\varphi)$. This suggests that the fluid dynamics of membrane is closely related to the dynamics of scalar field $\theta_{CS}$. To clarify this relation, more investigation should be done in the future by taking the potential of $\theta_{CS}$ into consideration.

Another or perhaps a more interesting result is about the Hall viscosity. Our result \eqref{Hav2} is much more complicated than that of reference \cite{Saremi:2011ab}, probably because we are studying a different spacetime. Since the calculation is too tedious, we will present the details and a careful comparison in a successive paper \cite{wang:2016}.

In this paper, we have focused mainly on asymptotically flat spacetimes. For the case with a nonvanishing cosmological constant, this paradigm has been considered very recently in various dimensions by references \cite{Fischler:2015kro,Landsteiner:2016stv} in the context of the AdS/CFT correspondence.

\begin{acknowledgments}
This work is supported by the Science and Technology Commission of Shanghai Municipality (Grant No. 11DZ2260700), and in part by the National Natural Science Foundation of China (Grant Nos. 11105053 and 91536218).
\end{acknowledgments}

\appendix


\section{Derivation of equation (\ref{spc})}\label{app-Cnh}
In this appendix, we will compute $C^{cb}n_{b}h_{c}^{~a}$ for the line element \eqref{metric} and eventually obtain equation \eqref{spc}.

Due to the Bianchi identity $\nabla_{c}R^{fb}_{~~de}+\nabla_{d}R^{fb}_{~~ec}+\nabla_{e}R^{fb}_{~~cd}=0$ and its contracted form $\nabla_{f}R^{fb}_{~~de}=\nabla_{d}R^{b}_{~e}-\nabla_{e}R^{b}_{~d}$, the Cotton tensor \eqref{cotton} can be reorganized into
\begin{eqnarray}
\nonumber C^{ab}&=&\frac{1}{2}\left[(\nabla_{c}\theta_{CS})\epsilon^{cdea}\nabla_{e}R^{b}_{~d}+(\nabla_{f}\nabla_{c}\theta_{CS})^{*}R^{cabf}\right]+(a\leftrightarrow b)\\
&=&\frac{1}{4}[(\nabla_{c}\theta_{CS})\nabla_{f}(\epsilon^{cade}R^{fb}_{~~de})-(\nabla_{f}\nabla_{c}\theta_{CS})\epsilon^{cade}R^{fb}_{~~de}]+(a\leftrightarrow b).
\end{eqnarray}
In order to work out its concrete form, we switch indices of equation \eqref{epsR} to obtain
\begin{eqnarray}
\nonumber\epsilon^{cade}R^{fb}_{~~de}&=&2\lambda_5\epsilon^{cade}(u^{f}u_{d}n^{b}n_{e}-u^{b}u_{d}n^{f}n_{e})+2\lambda_6\epsilon^{cade}(u^{f}u_{d}g^{b}_{~e}-u^{b}u_{d}g^{f}_{~e})\\
&&+2\lambda_7\epsilon^{cade}(n^{f}n_{d}g^{b}_{~e}-n^{b}n_{d}g^{f}_{~e})+2\lambda_8\epsilon^{cade}g^{f}_{~d}g^{b}_{~e}
\end{eqnarray}
and its covariant derivative
\begin{eqnarray}
\nonumber\nabla_{f}(\epsilon^{cade}R^{fb}_{~~de})&=&2\epsilon^{cade}(\nabla_{f}\lambda_{5})(u^{f}u_{d}n^{b}n_{e}-u^{b}u_{d}n^{f}n_{e})\\
\nonumber&&+2\epsilon^{cade}\lambda_{5}[-\lambda_{4}n_{d}n^{b}n_{e}-\lambda_{1}u_{d}u^{b}n_{e}+\lambda_{3}u_{d}n_{e}u^{b}\\
\nonumber&&-\lambda_{1}u_{d}u_{e}n^{b}+\lambda_{3}u_{d}u_{e}n^{b}-u^{b}u_{d}n_{e}(\nabla_{f}n^{f})-\lambda_{2}u^{b}u_{d}n_{e}-\lambda_{3}u^{b}u_{d}n_{e}]\\
\nonumber&&+2\epsilon^{cade}(\nabla_{f}\lambda_{6})(u^{f}u_{d}g^{b}_{~e}-u^{b}u_{d}g^{f}_{~e})\\
\nonumber&&+2\epsilon^{cade}\lambda_{6}[-\lambda_{4}n_{d}g^{b}_{~e}-\lambda_{4}u_{e}u_{d}n^{b}-\lambda_{4}u^{b}n_{d}u_{e}]\\
\nonumber&&+2\epsilon^{cade}(\nabla_{f}\lambda_{7})(n^{f}n_{d}g^{b}_{~e}-n^{b}n_{d}g^{f}_{~e})\\
\nonumber&&+2\epsilon^{cade}\lambda_{7}[(\nabla_{f}n^{f})n_{d}g^{b}_{~e}+\lambda_{2}n_{d}g^{b}_{~e}+\lambda_{3}n_{d}g^{b}_{~e}\\
\nonumber&&-\lambda_{1}u_{e}u^{b}n_{d}-\lambda_{2}n^{b}n_{d}n_{e}-\lambda_{3}n_{d}g^{b}_{~e}-\lambda_{1}u_{e}u_{d}n^{b}-\lambda_{2}n^{b}n_{d}n_{e}-\lambda_{3}n^{b}g_{de}]\\
&&+2\epsilon^{cade}(\nabla_{f}\lambda_{8})g^{f}_{~d}g^{b}_{~e}.
\end{eqnarray}

Now it is possible to map $C^{ab}$ to the $n_{a}h_{b}^{~i}$ direction. After lengthy and careful computation, we find
\begin{eqnarray}
\nonumber C^{ab}n_{a}h_{b}^{~i}&=&\frac{1}{2}(\nabla_{c}\theta_{CS})\epsilon^{cade}n_{a}[(\nabla_{f}\lambda_{6})u^{f}u_{d}h_{e}^{~i}-(\nabla_{f}\lambda_{6})u^{i}u_{d}g^{f}_{~e}+2(\nabla_{f}\lambda_{8})g^{f}_{~d}h_{e}^{~i}]\\
\nonumber&&+\frac{1}{2}(\nabla_{c}\theta_{CS})\epsilon^{cbde}h_{b}^{~i}[(\nabla_{f}\lambda_{5})u^{f}u_{d}n_{e}+(\nabla_{f}\lambda_{6})u^{f}u_{d}n_{e}+(\nabla_{f}\lambda_{8})g^{f}_{~d}n_{e}]\\
\nonumber&&-\frac{1}{2}(\nabla_{f}\nabla_{c}\theta_{CS})\epsilon^{cade}n_{a}(\lambda_{6}u^{f}u_{d}h_{e}^{~i}-\lambda_{6}u^{i}u_{d}g^{f}_{~e}+\lambda_{8}g^{f}_{~d}h_{e}^{~i})\\
&&-\frac{1}{2}(\nabla_{f}\nabla_{c}\theta_{CS})\epsilon^{cbde}h_{b}^{~i}(\lambda_{5}u^{f}u_{d}n_{e}+\lambda_{6}u^{f}u_{d}n_{e}-\lambda_{7}n_{d}g^{f}_{~e}+\lambda_{8}g^{f}_{~d}n_{e}).
\end{eqnarray}
Since $\lambda_{i}$'s are functions dependent only of $r$, with the help of equations \eqref{metric} and \eqref{nu}, one can check
\begin{equation}\label{hnlam}
u^{a}\nabla_{a}\lambda_{i}=h_{b}^{~a}\nabla_{a}\lambda_{i}=0.
\end{equation}
Therefore, it is straightforward to verify that
\begin{eqnarray}
\nonumber C^{ab}n_{a}h_{b}^{~i}&=&\frac{1}{2}(\nabla_{c}\theta_{CS})\epsilon^{cade}n_{a}[-(\nabla_{f}\lambda_{6})u^{i}u_{d}n^{f}n_{e}+2(\nabla_{f}\lambda_{8})n^{f}n_{d}h_{e}^{~i}]\\
\nonumber&&+\frac{1}{2}(\nabla_{c}\theta_{CS})\epsilon^{cbde}h_{b}^{~i}(\nabla_{f}\lambda_{8})n^{f}n_{d}n_{e}\\
\nonumber&&-\frac{1}{2}(\nabla_{f}\nabla_{c}\theta_{CS})\epsilon^{cade}n_{a}\lambda_{6}u^{f}u_{d}h_{e}^{~i}-\frac{1}{2}(\nabla_{f}\nabla_{c}\theta_{CS})n_{a}(-\lambda_{6}u^{i}u_{d}\epsilon^{cadf}+\lambda_{8}\epsilon^{cafe}h_{e}^{~i})\\
\nonumber&&-\frac{1}{2}(\nabla_{f}\nabla_{c}\theta_{CS})\epsilon^{cbde}h_{b}^{~i}(\lambda_{5}u^{f}u_{d}n_{e}+\lambda_{6}u^{f}u_{d}n_{e})\\
&&-\frac{1}{2}(\nabla_{f}\nabla_{c}\theta_{CS})h_{b}^{~i}(-\lambda_{7}n_{d}\epsilon^{cbdf}+\lambda_{8}\epsilon^{cbfe}n_{e}).
\end{eqnarray}
Because $\epsilon^{cade}$ is a completely antisymmetric tensor, while $\nabla_{f}\nabla_{b}\theta_{CS}$ is symmetric with respect to indices $f$ and $b$, the above equation is significantly simplified,
\begin{eqnarray}
\nonumber C^{ab}n_{a}h_{b}^{~i}&=&-\frac{1}{2}(\nabla_{f}\nabla_{c}\theta_{CS})\epsilon^{cade}n_{a}\lambda_{6}u^{f}u_{d}h_{e}^{~i}-\frac{1}{2}(\nabla_{f}\nabla_{c}\theta_{CS})\epsilon^{cbde}h_{b}^{~i}(\lambda_{5}u^{f}u_{d}n_{e}+\lambda_{6}u^{f}u_{d}n_{e})\\
\nonumber&=&-\frac{1}{2}(\nabla_{f}\nabla_{c}\theta_{CS})\epsilon^{cbde}(n_{b}\lambda_{6}u^{f}u_{d}h_{e}^{~i}+h_{b}^{~i}\lambda_{6}u^{f}u_{d}n_{e})-\frac{1}{2}(\nabla_{f}\nabla_{c}\theta_{CS})\epsilon^{cbde}h_{b}^{~i}\lambda_{5}u^{f}u_{d}n_{e}\\
&=&-\frac{1}{2}(\nabla_{f}\nabla_{c}\theta_{CS})\epsilon^{cide}\lambda_{5}u^{f}u_{d}n_{e}.
\end{eqnarray}
Keeping in mind equation \eqref{hnlam} and what is more,
\begin{eqnarray}
\nonumber\epsilon^{daec}u^{b}u_{e}n_{c}\nabla_{b}\nabla_{d}\lambda_{i}&=&\epsilon^{daec}u^{b}u_{e}n_{c}\nabla_{b}[(h_{d}^{~f}+n_{d}n^{f})\nabla_{f}\lambda_{i}]\\
\nonumber&=&\epsilon^{daec}u^{b}u_{e}n_{c}[n_{d}\nabla_{b}(n^{f}\nabla_{f}\lambda_{i})+(\nabla_{b}n_{d})n^{f}\nabla_{f}\lambda_{i}]\\
\nonumber&=&\epsilon^{daec}u^{b}u_{e}n_{c}(\lambda_1u_{b}u_{d}+\lambda_2n_{b}n_{d}+\lambda_3g_{bd})n^{f}\nabla_{f}\lambda_{i}\\
&=&0,
\end{eqnarray}
we can finally get
\begin{eqnarray}
\nonumber C^{cb}n_{b}h_{c}^{~a}&=&-\frac{1}{2}(\nabla_{f}\nabla_{c}\theta_{CS})\epsilon^{cade}\lambda_{5}u^{f}u_{d}n_{e}\\
\nonumber&=&-\frac{1}{2}\epsilon^{cade}u^{f}u_{d}n_{e}[\nabla_{f}\nabla_{c}(\lambda_{5}\theta_{CS})-\theta_{CS}\nabla_{f}\nabla_{c}\lambda_{5}-(\nabla_{f}\lambda_{5})\nabla_{c}\theta_{CS}-(\nabla_{f}\theta_{CS})\nabla_{c}\lambda_{5}]\\
&=&-\frac{1}{2}\epsilon^{cade}u^{f}u_{d}n_{e}\nabla_{f}\nabla_{c}(\lambda_{5}\theta_{CS}).
\end{eqnarray}


\begin{thebibliography}{99}
\bibitem{Damour:1978cg}
  T.~Damour,
  Phys.\ Rev.\ D {\bf 18}, 3598 (1978).

\bibitem{Damour79}
  T.~Damour,
  Ph.D. thesis,
  University of Paris VI, 1979.

\bibitem{Damour82}
  T.~Damour,
  in \emph{Proceedings of the Second Marcel Grossman Meeting on General Relativity},
  edited by R. Ruffini (North-Holland, Amsterdam, 1982), p. 587.

\bibitem{Price:1986yy}
  R.~H.~Price and K.~S.~Thorne,
  Phys.\ Rev.\ D {\bf 33}, 915 (1986).

\bibitem{Thorne86}
  K.~S.~Thorne, R.~H.~Price and D.~A.~Macdonald,
  ``\emph{Black Holes: The Membrane Paradigm},''
  Yale University Press, 1986.

\bibitem{Parikh:1997ma}
  M.~Parikh and F.~Wilczek,
  Phys.\ Rev.\ D {\bf 58}, 064011 (1998)
  [gr-qc/9712077].

\bibitem{Ashtekar:2000hw}
  A.~Ashtekar, S.~Fairhurst and B.~Krishnan,
  Phys.\ Rev.\ D {\bf 62}, 104025 (2000)
  [gr-qc/0005083].

\bibitem{Booth:2001gx}
  I.~S.~Booth,
  Class.\ Quant.\ Grav.\  {\bf 18}, 4239 (2001)
  [gr-qc/0105009].

\bibitem{Padmanabhan:2010rp}
  T.~Padmanabhan,
  Phys.\ Rev.\ D {\bf 83}, 044048 (2011)
  [arXiv:1012.0119 [gr-qc]].

\bibitem{Jaramillo:2013rda}
  J.~L.~Jaramillo,
  Phys.\ Rev.\ D {\bf 89}, 021502 (2014)
  [arXiv:1309.6593 [gr-qc]].

\bibitem{Chatterjee:2010gp}
  S.~Chatterjee, M.~Parikh and S.~Sarkar,
  Class.\ Quant.\ Grav.\  {\bf 29}, 035014 (2012)
  [arXiv:1012.6040 [hep-th]].

\bibitem{Jacobson:2011dz}
  T.~Jacobson, A.~Mohd and S.~Sarkar,
  arXiv:1107.1260 [gr-qc].

\bibitem{Kolekar:2011gg}
  S.~Kolekar and D.~Kothawala,
  JHEP {\bf 1202}, 006 (2012)
  [arXiv:1111.1242 [gr-qc]].

\bibitem{Wang:2014aty}
  T.~Wang,
  Class.\ Quant.\ Grav.\  {\bf 32}, no. 19, 195006 (2015)
  [arXiv:1411.6445 [gr-qc]].

\bibitem{Jackiw:2003pm}
  R.~Jackiw and S.~Y.~Pi,
  Phys.\ Rev.\ D {\bf 68}, 104012 (2003)
  [gr-qc/0308071].

\bibitem{Alexander:2009tp}
  S.~Alexander and N.~Yunes,
  Phys.\ Rept.\  {\bf 480}, 1 (2009)
  [arXiv:0907.2562 [hep-th]].

\bibitem{Grumiller:2007rv}
  D.~Grumiller and N.~Yunes,
  Phys.\ Rev.\ D {\bf 77}, 044015 (2008)
  [arXiv:0711.1868 [gr-qc]].

\bibitem{Lemos:2011ic}
  J.~P.~S.~Lemos and O.~B.~Zaslavskii,
  Phys.\ Rev.\ D {\bf 84}, 064017 (2011)
  [arXiv:1108.1801 [gr-qc]].

\bibitem{Delsate:2014hba}
  T.~Delsate, D.~Hilditch and H.~Witek,
  Phys.\ Rev.\ D {\bf 91}, no. 2, 024027 (2015)
  [arXiv:1407.6727 [gr-qc]].

\bibitem{Saremi:2011ab}
  O.~Saremi and D.~T.~Son,
  JHEP {\bf 1204}, 091 (2012)
  [arXiv:1103.4851 [hep-th]].

\bibitem{Jensen:2011xb}
  K.~Jensen, M.~Kaminski, P.~Kovtun, R.~Meyer, A.~Ritz and A.~Yarom,
  JHEP {\bf 1205}, 102 (2012)
  [arXiv:1112.4498 [hep-th]].

\bibitem{wang:2016}
T. Wang, in preparation.

\bibitem{Grumiller:2008ie}
  D.~Grumiller, R.~B.~Mann and R.~McNees,
  Phys.\ Rev.\ D {\bf 78}, 081502 (2008)
  [arXiv:0803.1485 [gr-qc]].

\bibitem{York:1972sj}
  J.~W.~York, Jr.,
  Phys.\ Rev.\ Lett.\  {\bf 28}, 1082 (1972).

\bibitem{Gibbons:1976ue}
  G.~W.~Gibbons and S.~W.~Hawking,
  Phys.\ Rev.\ D {\bf 15}, 2752 (1977).

\bibitem{Mann:2005yr}
  R. ~B. ~Mann and D. ~Marolf,
  Class. \ Quant. \ Grav. \  {\bf 23}, 2927 (2006)
  [hep-th/0511096].

\bibitem{Gourgoulhon:2005ng}
  E.~Gourgoulhon and J.~L.~Jaramillo,
  Phys.\ Rept.\  {\bf 423}, 159 (2006)
  [gr-qc/0503113].

\bibitem{Gourgoulhon:2005ch}
  E.~Gourgoulhon,
  Phys.\ Rev.\ D {\bf 72}, 104007 (2005)
  [gr-qc/0508003].

\bibitem{Fischler:2015kro}
  W.~Fischler and S.~Kundu,
  JHEP {\bf 1604}, 112 (2016)
  [arXiv:1512.01238 [hep-th]].

\bibitem{Landsteiner:2016stv}
  K.~Landsteiner, Y.~Liu and Y.~W.~Sun,
  arXiv:1604.01346 [hep-th].

\end{thebibliography}
 \end{document}